\documentclass[5p]{elsarticle}
\newcommand{\Fermi}{\textit{Fermi} }

\def\units#1{\hbox{$\,{\rm #1}$}}
\def\degrees{\hbox{${}^\circ$}}

\begin{document}

\title{Spectral analysis of the Crab Pulsar and Nebula with the {\Fermi}
Large Area Telescope}

\author[bari]{F.~Loparco\corref{cor1}}
\ead{francesco.loparco@ba.infn.it}

\cortext[cor1]{on behalf of the {\Fermi} LAT Collaboration}

\address[bari]{Universit\`a degli Studi di Bari e INFN Sezione di
Bari, Via E.~Orabona 4, 70126 Bari (Italy)}

\begin{keyword}
Crab Pulsar \sep Crab Nebula

\PACS 11.xx.22
\end{keyword}
  
\begin{abstract}

The Crab Pulsar is a relatively young neutron star. The pulsar is the
central star in the Crab Nebula, a remnant of the supernova SN 1054, which
was observed on Earth in the year 1054. The Crab Pulsar has been extensively
observed in the gamma-ray energy band by the Large Area Telescope
(LAT), the main instrument onboard the \Fermi Gamma-ray Space
Telescope, during its first months of data taking. The LAT data 
have been used to reconstruct the fluxes and the energy spectra
of the pulsed gamma-ray component and of the gamma-rays from the nebula.
The results on the pulsed component are in good agreement with the
previous measurement from EGRET, while the results on the nebula are
consistent with the observations from Earth based telescopes.

\end{abstract}

\maketitle

\section{Introduction}

The \Fermi Gamma-Ray Space Telescope was launched by NASA on June 11,
2008 from Cape Canaveral, and started taking data in sky survey
mode and in nominal science configuration on August 4, 2008. 
The main instrument onboard \Fermi is the Large Area Telescope 
(LAT)~\cite{Fermi}, a pair conversion telescope that covers a 
photon energy range extending from $20\units{MeV}$ beyond $300\units{GeV}$. 
\Fermi is moving on a quasi-circular orbit at 
$565 \units{km}$ altitude with an inclination of $25.6\degrees$,
and observes the whole sky every 2 orbits (corresponding 
to $\sim 3 \units{hours}$).

The Crab Pulsar and nebula are the remnants of the explosion
of a supernova reported by Chinese astronomers in 1054.
The Crab Pulsar, with a spin-down power 
$\dot{E} = 4.6 \cdot 10^{38} \units{erg~s^{-2}}$, 
is one of the most energetic
known pulsars. It is located at a distance of about $2\units{kpc}$
~\cite{Trimble} from Earth and is spinning with a period 
of $33\units{ms}$~\cite{Lyne}. 

Both the Pulsar and Nebula have been extensively observed 
in all the energy bands, from radio to high energy $\gamma$-rays.
In this paper we report the results of a spectral analysis
performed on both the Crab Pulsar and Nebula using the
data collected by the \Fermi LAT during its first 8 months 
of operation.

\begin{figure}[!ht]
\begin{center}
\includegraphics[width=0.9\linewidth]{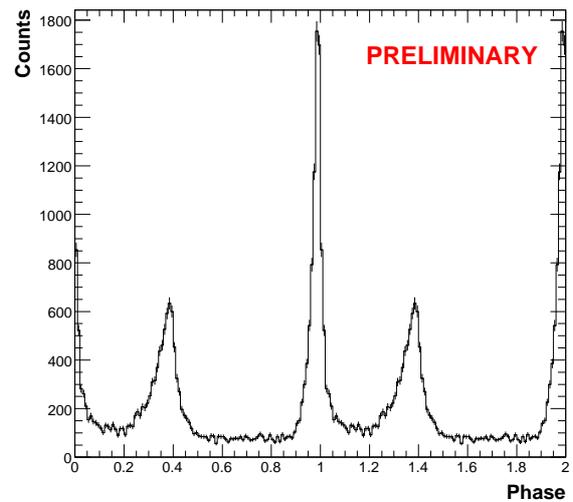}   
\end{center}
\caption{Phase histogram of the Crab Pulsar obtained
with photons above $100 \units{MeV}$. The main 
radio peak is at phase 0. Two cycles are shown.} 
\label{fig:phasogram}
\end{figure}

\section{Radio timing analysis}

The timing solution for the Crab Pulsar is built
using the data collected from the Nan\c cay and Jodrell 
radiotelescopes, that are involved in the LAT
pulsar timing campaign~\cite{Smith}. 
The Crab ephemeris has been built with the 
TEMPO2~\cite{Tempo2} package, using a set of $488$ 
observations at $600 \units{MHz}$ and $210$ observations 
at $1.4 \units{GHz}$, collected between August 2008 and 
April 2009~\cite{Grondin,Crab}.

\begin{figure*}[!ht]
\begin{center}
\includegraphics[width=0.9\linewidth]{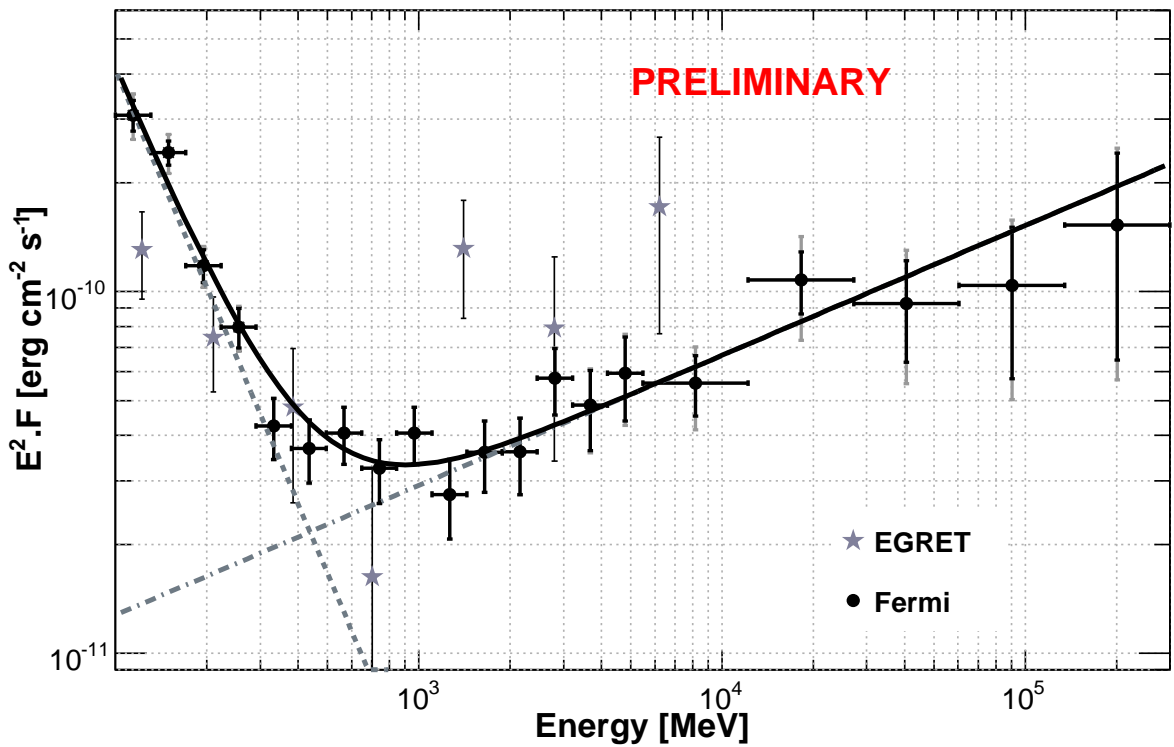}   
\end{center}
\caption{Spectral energy distribution of the Crab Nebula 
in the energy range from $100 \units{MeV}$ to $300 \units{GeV}$.
The black continuous line is the fitted spectrum of 
equation~\ref{eq:nebula}. The dotted and dotted-dashed lines 
represent respectively the synchrotron and the Inverse Compton 
components of the spectrum. The spectral points have been obtained 
performing maximum likelihood fits in individual energy
bins. The black error bars are statistical errors, while the grey
error bars represent statistical errors added in quadrature to 
systematic errors. The stars are the EGRET results, and
have been taken from ref.~\cite{egretneb}.} 
\label{fig:nebula}
\end{figure*}

\begin{figure*}[!ht]
\begin{center}
\includegraphics[width=0.9\linewidth]{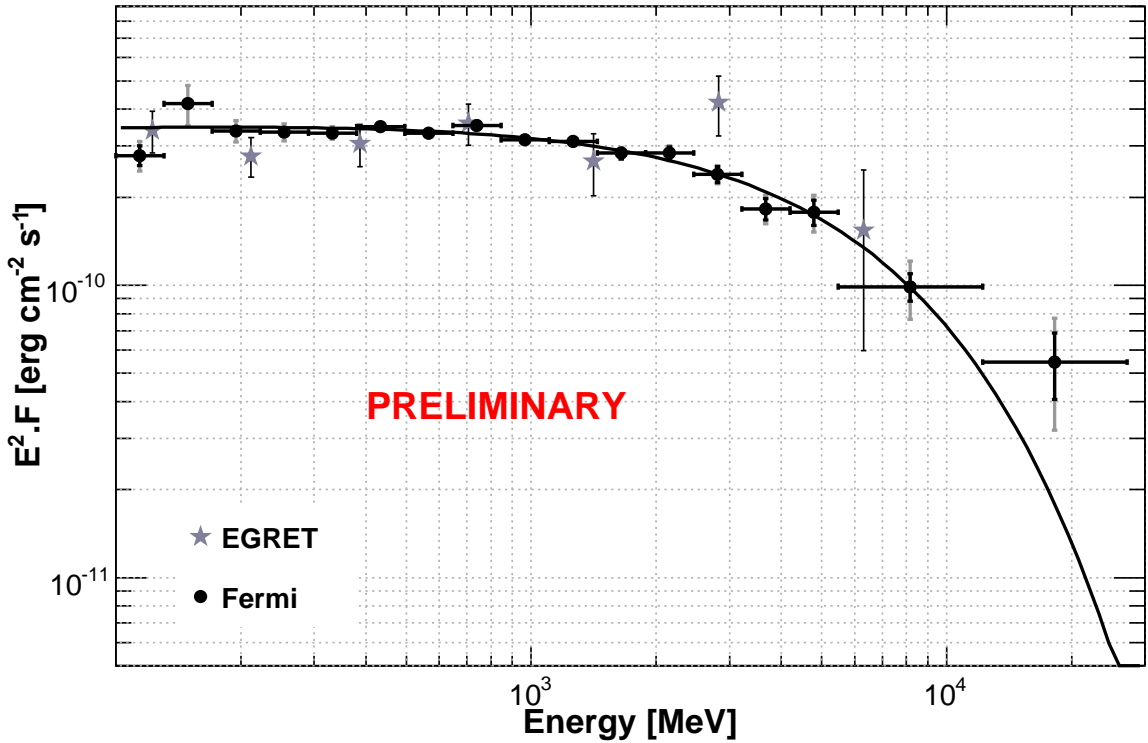}   
\end{center}
\caption{Spectral energy distribution of the Crab Pulsar 
in the energy range from $100 \units{MeV}$ to $300 \units{GeV}$.
The black continuous line is the fitted spectrum of 
equation~\ref{eq:pulsar}. The spectral points have been obtained 
performing maximum likelihood fits in individual energy
bins. The black error bars are statistical errors, while the grey
error bars represent statistical errors added in quadrature to 
systematic errors. The stars are the EGRET results, and
have been taken from ref.~\cite{egretneb}.} 
\label{fig:pulsar}
\end{figure*}

\section{Analysis of the \Fermi LAT data}

\subsection{Event selection}
\label{sec:evsel}

For this analysis~\cite{Grondin,Crab} we have considered 
a data sample
covering the period from August 2, 2008 to April 7, 2009. 
In particular, we have used only ``diffuse'' class~\cite{Fermi} 
photon events, that ensure the most effective background
rejection. Events with zenith angles larger than
$105\degrees$ have been excluded from the analysis, in
order to avoid contaminations from the Earth albedo
gamma-rays.

Figure~\ref{fig:phasogram} shows the Crab
Pulsar phase histogram obtained selecting photons
with energies above $100 \units{MeV}$ and with an 
angular separation from the nominal source position
less than
\begin{displaymath}
\theta_{max} = max \left\{ \left[ 6.68\degrees - 1.76\degrees \log_{10} E
(MeV) \right] , 1.3\degrees \right\}
\end{displaymath}
where $E$ is the photon energy. The phase $0$ corresponds
to the position of the main radio peak.
The Crab phase histogram exhibits two peaks
at phases $\phi_{1}=0.9915 \pm 0.0005$ and 
$\phi_{2}=0.3894 \pm 0.0022$ respectively. The off-pulse 
window is defined in the phase range from $0.52$ to
$0.87$.

\subsection{Spectral analysis of the Nebula}
\label{sec:nebula}

The spectral analysis has been performed following
two different and complementary strategies. 
In the first case, the spectra are reconstructed using 
a maximum likelihood approach, that is implemented 
in the \Fermi science tool ``{\em gtlike}''~\cite{gtlike}.
In the second case, the spectra have been 
evaluated using an iterative unfolding procedure based on
the Bayes theorem~\cite{unfolding,Mazziotta}.

The maximum likelihood spectral analysis of the nebula
has been performed selecting photons in the
off-pulse window, in a $20\degrees$ region 
centered on the nominal pulsar position, and
with energies in the range from 
$100\units{MeV}$ to $300\units{GeV}$. 
The galactic background component has been modeled
using GALPROP~\cite{galprop}, while the extragalactic component
and the instrumental background have been described
with a single isotropical power law spectrum.
The Crab Nebula spectrum is well described by
a sum of two power law spectra:
\begin{equation}
\label{eq:nebula}
\frac{dN}{dE} = k_{S} \left(\frac{E}{E_{0}}\right)^{-\Gamma_{S}} 
	     + k_{IC} \left(\frac{E}{E_{0}}\right)^{-\Gamma_{IC}}
\end{equation}
where $E_{0}=1\units{GeV}$.
The two terms at the right hand side of eq.~\ref{eq:nebula}
are labeled with the indices $S$ and $IC$ because they 
are identified respectively with the falling edge of the 
synchrotron component and with the rising edge of the
Inverse Compton component of the spectrum.

The best fit values of the parameters are
$k_{S}= (9.1 \pm 2.1 \pm 0.7) \units{cm^{-2}s^{-1}MeV^{-1}}$,
$\Gamma_{S}=3.99 \pm 0.12 \pm 0.08$,
$k_{IC}= (6.4 \pm 0.7 \pm 0.1) \units{cm^{-2}s^{-1}MeV^{-1}}$
and $\Gamma_{IC}=1.64 \pm 0.05 \pm 0.07$, where both
statistical and systematic errors are quoted.
The latter, that are originated from the 
uncertainties in the LAT response,  have been 
evaluated using modified instrument response 
functions with respect to the nominal ones.

Figure~\ref{fig:nebula} shows the spectral
energy distribution $E^{2}dN/dE$ of the Crab Nebula. 
The spectral points superimposed to best fit function 
have been obtained performing a maximum likelihood 
fit in individual energy bins. In this case 
a simple power law shape has been assumed for the
source spectrum in each energy bin. 
In figure~\ref{fig:nebula} the \Fermi results are 
also compared with the spectral point obtained
by EGRET~\cite{egretneb}. At energies below $1 \units{GeV}$
the flux measured by \Fermi is higher than the one 
measured by EGRET. On the other hand, at higher 
energies, EGRET measured a higher flux than \Fermi.

In the unfolding analysis, the Crab Nebula spectrum has
been reconstructed without assuming any spectral model and
without performing any fit. This analysis technique also
allows to take into account the energy dispersion 
introduced by the detector response, that is not 
considered in the maximum likelihood approach.
The observed count spectrum of the Nebula, used as input for
the unfolding analysis, is built selecting the photons in the
off-pulse window, and subtracting the background counts, 
evaluated using the background model obtained
from the ``gtlike'' fit.
The unfolded spectrum of the Nebula is in good agreement with 
the one obtained from the likelihood analysis.

\subsection{Spectral analysis of the Pulsar}
\label{sec:pulsar}

The spectral analysis of the Crab Pulsar has been performed
with the same two techniques used for the Nebula.

The maximum likelihood analysis of the pulsed
emission has been done over the whole phase interval, 
including the Nebula in the background. The values of 
the spectral parameters of the Nebula reported in
section~\ref{sec:nebula} have been renormalized to
match the whole phase interval and have been fixed.
The pulsar spectrum has been fitted with an 
exponential cutoff power law spectrum given by:
\begin{equation}
\label{eq:pulsar}
\frac{dN}{dE} = k \left(\frac{E}{E_{0}}\right)^{-\Gamma} 
	\exp \left( -\frac{E}{E_{c}} \right)
\end{equation}
where $E_{0}=1\units{GeV}$.

The best fit values of the parameters are
$k = (2.36 \pm 0.06 \pm 0.15) \units{cm^{-2}s^{-1}MeV^{-1}}$,
$\Gamma = 1.97 \pm 0.02 \pm 0.06$,
and $E_{c}=(5.8 \pm 0.5 \pm 1.2) \units{GeV}$, where both
statistical and systematic errors are quoted.
We have also tried to perform a fit using a
power law with a super-exponential cutoff, but 
the likelihood value was not significantly 
better than the one obtained using the
function in eq.~\ref{eq:pulsar}.

Figure~\ref{fig:pulsar} shows the spectral
energy distribution $E^{2}dN/dE$ of the Crab Pulsar. 
The spectral points superimposed to best fit function 
have been obtained performing a maximum likelihood 
fit in individual energy bins, with the same
procedure described in section~\ref{sec:nebula}. 
In figure~\ref{fig:pulsar} the \Fermi results are 
also compared with the spectral point obtained
by EGRET~\cite{egretneb}. The \Fermi results are
consistent with the EGRET ones in the energy range 
up to $8 \units{GeV}$.

The spectrum of the Crab Pulsar has been also
reconstructed using the unfolding analysis.
In this case the observed count spectrum of the Pulsar
is built selecting photons in the on-pulse
phase window, and subtracting the background counts
in the off-pulse window, properly rescaled for
the on/off phase ratio (i.e. $0.65/0.35$ according
to the definition in section~\ref{sec:evsel}).
As for the Nebula, the unfolded spectrum of the Pulsar 
is consistent with the one evaluated from the maximum 
likelihood analysis.

\section{Conclusions}

We have measured the energy spectra of the 
Crab Pulsar and Nebula in the energy interval
from $100\units{MeV}$ to $300\units{GeV}$
using a sample of data taken
by the \Fermi LAT during its first 8 months of operation
in survey mode. 

The spectrum of the Nebula is well modeled with
a sum of two power laws, with spectral indices
$\Gamma_{S}=3.99 \pm 0.12 \pm 0.08$
and $\Gamma_{IC}=1.64 \pm 0.05 \pm 0.07$, 
which describe respectively the falling edge of the synchrotron
component and the rising edge of the Inverse Compton 
component. The pulsed emission
is described by a power law spectrum with an exponential
cutoff with spectral index
$\Gamma = 1.97 \pm 0.02 \pm 0.06$
and cutoff energy $E_{c}=(5.8 \pm 0.5 \pm 1.2) \units{GeV}$.

The spectral analysis has been performed 
using two different and complementary strategies, 
based respectively on a maximum likelihood 
fit and on an unfolding analysis. The spectra
reconstructed with these two analysis techniques
are in good agreement, thus confirming the
reliability of our measurement.

The \Fermi LAT has allowed a precise
measurement of the spectra of both the
Crab Pulsar and Nebula that has been made possible by
the wider energy range and by the better 
energy sensitivity with respect to its
predecessors.

\section*{Acknowledgements}

The \Fermi LAT Collaboration acknowledges support from a number of
agencies and institutes for both development and the operation of 
the LAT as well as scientific data analysis. These include NASA and
DOE in the United States, CEA/Irfu and IN2P3/CNRS in France, ASI 
and INFN in Italy, MEXT, KEK, and JAXA in Japan, and the 
K.~A.~Wallenberg Foundation, the Swedish Research Council and 
the National Space Board in Sweden. Additional support from INAF 
in Italy for science analysis during the operations phase is also 
gratefully acknowledged.

The Nan\c cay Radio Observatory is operated by the Paris Observatory, 
associated with the French Centre National de la Recherche Scientifique (CNRS).

The Lovell Telescope is owned and operated by the University of
Manchester as part of the Jodrell Bank Centre for Astrophysics with 
support from the Science and Technology Facilities Council of the United Kingdom.

\end{document}